# Quantum Computing with Neutral Atoms in an Optical Lattice

Ivan H. Deutsch and Gavin K. Brennen

*Dept. of Physics and Astronomy, University of New Mexico, Albuquerque, NM 87131*

Poul S. Jessen

*Optical Sciences Center, University of Arizona, Tucson, AZ 85721*

**Abstract:** We present a proposal for quantum information processing with neutral atoms trapped in optical lattices as qubits. Initialization and coherent control of single qubits can be achieved with standard laser cooling and spectroscopic techniques. We consider entangling two-qubit logic gates based on optically induced dipole-dipole interactions, calculating a figure-of-merit for various protocols. Massive parallelism intrinsic to the lattice geometry makes this an intriguing system for scalable, fault-tolerant quantum computation.

## I. The Tao of Quantum Computing

The end of the twentieth century has seen a remarkable synergy of two of its most important scientific achievements: information science and quantum physics. Following the pioneering work of Benioff [1], Feynmann [2], Deutsch [3], Bennett [4], Shor [5] and others, we are just beginning to appreciate the tremendous potential of devices that represent and process information according to the laws of quantum mechanics. Much attention is currently focussed on the idea of "quantum computation", according to which algorithms can be carried out as unitary transformations on a many-body quantum system [6], sometimes with enormous savings in computational resources compared to any classical device. Examples of the impact of quantum information science (QIS) in other areas include improved precision measurements [7], quantum simulations [8] and quantum communication [9].

While there has been rapid progress in developing the fundamental notions of QIS, it remains a grand challenge to bring its promises to fruition in the laboratory. In the prevalent paradigm, a unitary transformation (computation) is performed by a quantum circuit consisting of a sequence of quantum gates acting upon a collection of qubits (two-level quantum systems). To allow this, a physical implementation of quantum computing must satisfy two requirements:

- The constituent particles must be *strongly* coupled to one another, to the external coherent fields which drive the unitary evolution, and to an external $T \approx 0K$ bath that provides for qubit initialization.
- The constituent particles must be *weakly* coupled to the incoherent noisy environment which leads to decoherence and ultimately reduces quantum-information to classical-information.

This inherent conflict, which we call the "Tao of Quantum Computing", highlights the daunting realities of laboratory experiments. The key challenge of QIS is then to find or engineer a quantum many-body system so that coherent evolution and decoherence can take place on very different time scales. There are of course further requirements for implementing a general purpose quantum computer, including input-output and error-correction, as discussed by DiVincenzo in the introductory chapter to this volume. The diverse proposals discussed within this special issue all make different compromises with these various demands in mind. For example in the ion trap geometry (see article by Wineland in this volume), strong coherent coupling is provided by the repulsive Coulomb interaction, but this interaction also couples strongly to the perturbing effects of the noisy environment [10]. A very different compromise is made by NMR (see Cory in this volume) which employs the much weaker magnetic dipole-dipole interactions between nuclei in molecules. This system has intrinsically long coherence times and weak coupling to the environment. However, these natural systems come at a price. Such weak interactions between nuclei occur only at very short distances in which case addressing and distinguishing qubits for a large collection becomes problematic. An alternative is to engineer an artificial system for which addressing and preparing qubits is built into the design (see articles by Kane and also Tanamoto). The decoherence properties for these systems are promising but have so far been measured only for non-interacting qubits. In accordance with the Tao of Quantum Computing one can expect that coupling qubits to each other will bring with it stronger coupling to the environment (e. g. the host silicon lattice) and faster decoherence. As such implementations of QIS move to the laboratory, it will be interesting to see what new compromises can be worked out for each case.

We have recently considered in some detail how to encode and process quantum information using trapped neutral atoms as qubits. This approach has several appealing features. Neutral atoms in their electronic ground state couple extremely weakly to the environment, allowing potentially very long coherence times [11]. In most situations neutrals also couple very weakly to each other, but interatomic couplings can be created *on demand* by induced electric dipole-dipole interactions [12, 13], ground state collisions [14, 15], or by real photon exchange [16]. The ability to turn interactions "on" only when needed is highly advantageous because it reduces coupling to the environment and the spread of errors during computation. The weak atomic interactions also make it relatively straightforward to trap and cool neutrals in large numbers, with favorable implications for scaling to many qubits and perhaps parallel processing.

We discuss in this paper several different protocols to carry out quantum logic operations, focusing on one possible entangling mechanism, laser induced electric dipole-dipole interactions. Quantum information is encoded in a pair of internal states in the atomic hyperfine ground manifold, and the excitation of a dipole moment is made conditional on the state of each atomic qubit. Radiative damping of the dipoles brings with it a fundamental mechanism of decoherence (the Tao of Quantum Computing), which must be suppressed by performing all manipulations rapidly compared to the rate of spontaneous emission. A simple scaling argument shows that one can in principle obtain the necessary separation of timescales. Consider two atoms separated at a distance $r \ll \lambda$, with an interaction energy that scales as $V_{dd} \sim \langle \mathbf{d} \rangle^2 / r^3$, where $\langle \mathbf{d} \rangle$ is the mean induced dipole moment per atom. The induced dipoles can spontaneously emit photons, but this process is bounded above by the Dicke-superradiance cooperative decay rate, which is equal to twice the single atom scattering rate $\Gamma' \sim k_L^3 \langle \mathbf{d} \rangle^2 / \hbar$, where $k_L$ is the wave number of the driving laser. We can characterize the performance of a quantum "gate" by a figure of merit that measures the ratio of the coherent interaction energy of two qubits to their collective decoherence rate, $\kappa = V_{dd} / \hbar \Gamma' \sim (k_L r)^{-3}$. For sufficiently tight atomic confinement, $k_L r \ll 1$ (the Lamb-Dicke regime), the dipole-dipole level shift can be much larger than the scattering rate and the interaction effectively becomes coherent.

At this time, the best prospects for such tight confinement is offered by a type of neutral atom trap known as an optical lattice [17]. Optical lattices are periodic arrays of micron-sized traps created by the ac-Stark shift in the interference pattern of a set of intersecting laser beams. Using lattice light detuned far from atomic resonance it is possible to greatly reduce the rate of photon scattering, while at the same time a large intensity allows us to maintain a strong trapping potential. Some of the building blocks of quantum computing have already been demonstrated in this system, including cooling to the vibrational ground state of the lattice microtraps [18], and coherent control of the combined center-of-mass and ground hyperfine degrees of freedom [19]. A wide range of properties characterizing the lattice potential can be controlled in real time through changes in laser beam geometry, polarization, intensity and frequency, and through the addition of static electric and magnetic fields. Of particular interest is the ability to design trapping potentials that are very nearly identical for the two hyperfine spin states used to encode an atomic qubit, thereby largely decoupling the internal degrees of freedom which store the qubits from the center-of-mass motion. It is even possible to encode distinct subsets of atomic qubits, trapped on lattice grids that can be moved independently of each other. This provides a mechanism to superimpose and couple any two qubits from the two grids, and additionally lends itself naturally to moving and operating on blocks of qubits in a highly parallel fashion, as required for quantum error correction [20].

The remainder of this chapter is organized as follows. In Sec. II we give an overview of the atom/optical lattice system, defining the qubits and discussing how to prepare, manipulate and measure their state. In Sec. III we discuss the fundamental physics associated with our proposed entangling mechanism – optically induced dipole-dipole interactions. In Sec. IV we outline protocols for a universal set of quantum logic gates, and estimate a figure of merit for each member. Strategies for measuring and minimizing the gate error-probability are discussed in Sec. V. Finally in Sec. VI we given an outlook for near-term studies based on existing technology and towards future prospects of this system for robust, fault-tolerant operation.

## II. The Alkali/Optical Lattice System

When we choose an atomic species and design a neutral atom trap for use in QIS we must pay attention to several requirements:

- The intrinsic decoherence rate of the trap must be low.
- The trap must provide confinement on a scale much smaller than the optical wavelength.
- The trap must be compatible with the encoding of quantum information in an atomic internal and/or motional degree of freedom.
- The interaction between atomic qubits must be precisely controlled and programmable.

Two major classes of neutral atom traps are routinely used in the laboratory: magnetic traps based on the interaction $-\mu \cdot \mathbf{B}$ between an atomic magnetic moment and a magnetic field, and optical traps based on the interaction $-\mathbf{d} \cdot \mathbf{E}/2$ between an induced electric dipole moment and a laser field. Magnetic traps provide very long coherence times, as evidenced by the remarkable coherent dynamics of Bose-Einstein Condensed atomic vapors [21], but do not lend themselves easily to encoding and controlling large numbers of atomic qubits. By contrast, optical traps are intrinsically dissipative at some level due to photon scattering, but this disadvantage is more than compensated by the tremendous flexibility with which one can design the trapping potential through choice of atomic species and optical field parameters. So far, we and others have focussed on optical lattices, whose central feature compared to other types of optical traps is the ease with which one can trap large numbers of atoms and move them physically with precise control so that qubits can be pairwise superimposed and quantum logic operations carried out between them.

In the large detuning, low saturation limit of interest here, the light shift (ac Stark shift) of an atomic ground state is readily calculated using perturbation theory. For a monochromatic field, $\mathbf{E}(\mathbf{x},t) = \text{Re}\left(E_0(\mathbf{x})\vec{\varepsilon}(\mathbf{x})e^{-i\omega_L t}\right)$, and atomic tensor polarizability $\tilde{\vec{\alpha}}$ one has the essentially classical expression,

$$U(\mathbf{x}) = -\frac{|E_0(\mathbf{x})|^2}{4}\vec{\varepsilon}^*(\mathbf{x}) \cdot \tilde{\vec{\alpha}} \cdot \vec{\varepsilon}(\mathbf{x}). \tag{1}$$

This equation indicates that the optical potential can be designed through judicious choice of the field amplitude $E_0(\mathbf{x})$, polarization $\vec{\varepsilon}(\mathbf{x})$, and atomic species whose internal state dictates the polarizability tensor $\tilde{\vec{\alpha}} = -\text{Re}\sum_e \mathbf{d}_{ge}\mathbf{d}_{eg}/\hbar(\Delta_{eg} + i\Gamma_e/2)$. Here $\mathbf{d}_{eg}$ is the dipole operator connecting states $e$ and $g$, $\Delta_{eg} = \omega_L - \omega_{eg}$ is the detuning from the corresponding transition frequency, and $\Gamma_e$ is the excited state natural linewidth. Dissipation in the potential is characterized by the imaginary part of $\tilde{\vec{\alpha}}$, and arises due to spontaneous photon scattering which occurs at a rate $\Gamma' = \sum_e s_{eg}\Gamma_e/2$, where $s_{eg} = \Omega_{eg}^2/(2\Delta_{eg})^2$ is the saturation parameter for a given Rabi frequency $\Omega_{eg} = \mathbf{d}_{eg} \cdot \mathbf{E}/\hbar$. Since the scattering rates scale as $|\mathbf{E}|^2/\Delta_{eg}^2$ and the trap depth scales as $|\mathbf{E}|^2/\Delta_{eg}$ (Eq. (1)), a trap using intense light detuned far from any resonance can provide substantial confinement with minimal dissipation.

In the remainder of this paper we restrict our discussion to alkali atoms, which have become the staple of the atomic physics and quantum optics community due to the ease with which they can be laser cooled and trapped. The common isotopes of Na, Rb and Cs have nuclear spin and hyperfine structure, with $nS_{1/2}(F = I \pm \frac{1}{2})$ ground states and $nP_{1/2}(F = I \pm \frac{1}{2})$, $nP_{3/2}(F = I \pm \frac{1}{2}, \frac{3}{2})$ excited states (the well known D1 and D2 resonance lines). To provide an idea of the appropriate physical scales, consider a Cs atom trapped near a node of a linearly polarized standing wave with a peak intensity of $500\,\text{W/cm}^2$ and tuned 50 GHz blue of (i. e. above) the D2 resonance. This configuration forms a very simple one-dimensional (1D) optical lattice, with a potential $U(x) = U_0 \cos^2(k_L x)$, where $k_L$ is the laser wavenumber and the maximum light shift at the standing wave anti-nodes is $U_0 \sim 10^4 E_R \sim 1\,\text{mK}$. The single-photon recoil energy $E_R = (\hbar k_L)^2/2m$, where $m$ is the atomic mass, is used as a natural energy scale for the problem. The atomic center-of-mass oscillation frequency along the standing wave is $\omega_{osc} \sim 200 E_R/\hbar \sim$ 500 kHz and the rms spread of the vibrational ground state is $x_0 \sim \lambda_L/100 \sim 10$ nm. In the following we will characterize atomic confinement by the Lamb-Dicke parameter $\eta \equiv k_L x_0$, which is here $\sim 0.07$. By comparison, the rate of photon scattering is $\Gamma' \sim 65\,\text{s}^{-1}$ and $\omega_{osc}/\Gamma' \sim 6 \times 10^3$, i. e. the vibrational level structure is highly resolved and the atom undergoes many thousand oscillations between spontaneous emission events. Different compromises between oscillation frequency/confinement and photon scattering rate can be made according to Eq. (1), limited chiefly by the lattice intensity as dictated by available laser power.

Numerous extensions of the above 1D lattice to two and three dimensions have been explored in both theory and experiment [17]. Typically, these lattices have been loaded with atoms through a combination of standard laser cooling in a magneto-optic trap, precooling in a near-resonance optical lattice, and adiabatic transfer to the far-off-resonance lattice [22]. Due to the low heating rate and tight confinement the atoms can be further cooled to the quantum mechanical vibration ground-state of the far-off-resonance lattice with near unit efficiency, using the method of resolved-sideband Raman cooling [23], originally developed for ion traps [24]. A major shortcoming of this approach is that the initial laser cooled sample is relatively dilute and the eventual atom distribution over available lattice sites is correspondingly sparse and random, with perhaps one atom per $\sim 100$ lattice sites. Very recently, experiments have succeeded in loading from much denser samples, demonstrating both near-unit occupation of the lattice sites [25] and resolved-sideband cooling at correspondingly high atomic density [26].

For our implementation of quantum logic in this system we consider a three dimensional (3D) optical lattice geometry such as the one shown in Fig. 1, consisting of blue detuned standing waves aligned along the three Cartesian axes. Interference effects can be eliminated by choosing slightly different optical frequencies for beams propagating along different axes, so that the total lattice potential becomes the simple sum of three 1D lattice potentials. The two lattices in the *x-y* plane are formed by linearly polarized standing waves and confine atoms in a square array of potential wells corresponding to "tubes" along the *z* axis. In the *z* direction, we allow for a variable angle $\theta$ between the counterpropagating linear polarizations (lin-$\theta$-lin configuration [27]). These beams produce a pair of $\sigma_+ - \sigma_-$ standing waves with antinodes separated by $\delta z = \lambda \tan^{-1}[(\tan\theta)/2]$. The resulting optical potential permits us to trap two distinguishable sets of atoms according to Eq. (1): those in internal states most light-shifted by $\sigma_+$ light, and those in states most light-shifted by $\sigma_-$ light. We refer to these as the "(±)-species". The ability to dynamically control the angle $\theta$ between polarizations, and thus vary the distance between the nodes of the interleaved standing waves, allows us to separate and move atoms of the (±)-species relative to each other – one of the central features of our proposal. Two atoms of different species, initially separated along the *z*-axis by a distance $\Delta z$, can be made to overlap by rotating $\theta$ by $2\pi \Delta z/\lambda$. If this angular rotation is made adiabatic with respect to the oscillation frequency, then the center of mass wave functions of the atoms are unchanged.

Once the atoms have been brought together they can be made to interact. We focus in this paper on the possibility of employing optically induced electric dipole-dipole interactions by applying an auxiliary laser pulse, referred to as the "catalysis field", for a time necessary to achieve the desired two-qubit logical operations (for an alternative coupling scheme based on elastic collisions in a closely related proposal, see Briegel *et al.* in this volume). After a logical operation, the atoms can be separated by further adiabatic rotation of the lattice polarizations so that they no longer interact. The catalysis field is assumed to be tuned closer to resonance than the lattice fields and induces stronger dipoles (though still with $s \ll 1$), so that the dipoles induced by the lattice fields can be neglected in their contribution to the atom-atom interactions.

To make this discussion more specific we must choose a proper set of logical basis states for our atomic qubits. To do so it is useful to express the optical potential, Eq. (1), in terms of its irreducible components. For alkali atoms excited near the D2 resonance line, at a detuning much larger than the hyperfine splitting but less than the fine structure splitting in the excited state, the light shift can be cast in the form [19]

$$\hat{U}_F(\mathbf{x}) = U_J(\mathbf{x}) + \mathbf{B}_{eff}(\mathbf{x}) \cdot \frac{\hat{\mathbf{F}}}{F}$$

$$U_J(\mathbf{x}) = \frac{2}{3} U_1 |\vec{\varepsilon}_L(\mathbf{x})|^2, \quad \mathbf{B}_{eff}(\mathbf{x}) = -\frac{i}{3} U_1 [\vec{\varepsilon}_L^*(\mathbf{x}) \times \vec{\varepsilon}_L(\mathbf{x})],$$

(2)

where $U_1$ is the single-beam light shift. Expressing the light shift in terms of a scalar potential $U_J(\mathbf{x})$ independent of *F* and $M_F$, and an effective magnetic field $\mathbf{B}_{eff}(\mathbf{x})$ serves as a powerful guide for encoding quantum information. For example, if the trap polarization is everywhere linear then $\vec{\varepsilon}_L(\mathbf{x})$ is real and $\mathbf{B}_{eff}(\mathbf{x}) \equiv 0$, and the light shift operator is simply a scalar potential. The preceding discussion shows that this situation is too restrictive for our purposes since we lose the ability to move atoms relative to one another via polarization rotation. Instead, we seek a finite $\mathbf{B}_{eff}$ and choose to encode a qubit in two hyperfine ground states whose the $g_F$-factors have equal magnitude (to within 10[-3]) and opposite sign, so that states $\left| F_\uparrow = I + \tfrac{1}{2}, M_F \right\rangle$ and $\left| F_\downarrow = I - \tfrac{1}{2}, -M_F \right\rangle$ have nearly identical light shifts. There are several reasons why two such states form a robust basis for an atomic qubit. First, fluctuations in the light shift and/or Zeeman shift from ambient real magnetic fields will be common mode, and cause minimal phase accumulation between the qubit states. Second, an individual qubit always remains spatially localized as the lattice polarization is changed – the entanglement created by quantum logic operations occurs solely between the *spins* of different atoms. There is never any entanglement within a single atom between its internal and external degrees of freedom. This is in contrast to the proposal of Jaksch *et al.* [14], whereby the logical-$|1\rangle$ and logical-$|0\rangle$ of a given atom follow different potentials, leading to the formation of nonlocal entangled "Schrödinger-cat" states of individual atoms. Such states could potentially be highly susceptible to phase errors caused by fluctuations or spatial inhomogeneities in the trapping potential and/or ambient magnetic fields.

With these considerations in mind we can define suitable computational basis sets for the $(\pm)$-species atoms in our optical lattice. For example, we can choose

$$|1\rangle_{\pm} = |F_{\uparrow}, M_F = \pm 1\rangle \otimes |\psi_1\rangle_{ext}, \quad |0\rangle_{\pm} = |F_{\downarrow}, M_F = \mp 1\rangle \otimes |\psi_0\rangle_{ext} \tag{3}$$

where $|\psi\rangle_{ext}$ are the motional quantum states. In this basis, rotations on the single qubit Bloch sphere can be performed using coherent Raman pulses and ac-Stark shifts, with the $(+)$ and $(-)$ species addressed separately via the polarization of the Raman pulses. Projective measurements of the qubit state can be performed with conventional fluorescence spectroscopy. The spectroscopic tools for these tasks are equivalent to those demonstrated in ion traps [28].

Based on these ideas we can envision how an ideal optical lattice quantum computer might function. The interleaving "+" and "–" lattices of Fig. 1 are occupied in a well defined pattern by single atomic qubits in pure internal and vibrational ground states. This initializes the system in accordance with the first requirement of DiVincenzo. With no manipulation to bring atoms together in the *x-y* plane, each tube of atoms along *z* represents a separate quantum register. Within each register two-qubit logic gates are operated by bringing targeted pairs together through polarization rotation, applying an appropriate catalysis pulse which induces a cooperative interaction, and then separating the atoms again. In the following section we discuss the fundamentals of the dipole-dipole entangling mechanism and then protocols for implementing two-qubit logic gates such as CPHASE or $\sqrt{\text{SWAP}}$. These, together with our ability to carry out arbitrary single qubit manipulations, constitute the requisite universal set of logic gates. Finally, read out is performed as a projective measurement on individual qubits.

## III. Dipole-Dipole Interactions

The dipole-dipole interaction at the heart of our proposed two-qubit logic gate depends on both the internal electronic states of the atoms, as dictated by the tensor nature of the interaction, and the external motional states, which determine the relative coordinate probability distribution of the dipoles. We consider a system of two atoms trapped in harmonic wells, interacting coherently with a classical field and with each other via the dipole-dipole interaction. Decoherence may occur via cooperative spontaneous emission. We seek expressions for the interaction matrix elements and the resulting selection rules (for details see [13]).

Consider two alkali atoms with nuclear spin $I$ and center of mass positions $\mathbf{r}_1$, $\mathbf{r}_2$, excited on the D2 transition $|S_{1/2}(F)\rangle \leftrightarrow |P_{3/2}(F')\rangle$, where $F$ and $F'$ belong to the ground and excited state hyperfine manifolds. The atoms interact with the vacuum field and a classical monochromatic laser field, detuned by amount $\Delta$, large compared to the excited-state hyperfine splitting. Since the quantum information will be stored in the ground-state hyperfine sublevels, it is important for us to include the complex internal structure in calculating the interaction matrix elements. After tracing over the vacuum modes in the usual Born-Markov and rotating wave approximations, one obtains the effective Hamiltonian for the atom-laser interaction, together with a dipole-dipole interaction between atoms [29],

$$H_{AL} = -\hbar\left(\Delta + i\frac{\Gamma}{2}\right)\left(\mathbf{D}_1^\dagger \cdot \mathbf{D}_1 + \mathbf{D}_2^\dagger \cdot \mathbf{D}_2\right) - \frac{\hbar\Omega}{2}\left(\mathbf{D}_1^\dagger \cdot \vec{\varepsilon}_L(\mathbf{r}_1) + \mathbf{D}_2^\dagger \cdot \vec{\varepsilon}_L(\mathbf{r}_2) + h.c.\right), \tag{4a}$$

$$H_{dd} = V_{dd} - i\frac{\hbar\Gamma_{dd}}{2} = -\frac{\hbar\Gamma}{2}\left(\mathbf{D}_2^\dagger \cdot \vec{\mathbf{T}}(k_L r) \cdot \mathbf{D}_1 + \mathbf{D}_1^\dagger \cdot \vec{\mathbf{T}}(k_L r) \cdot \mathbf{D}_2\right). \tag{4b}$$

The dimensionless dipole raising operator associated with absorption of a photon is defined as

$$\mathbf{D}^\dagger = \sum_{F'} \frac{P_{F'}\mathbf{d}P_F}{\langle J'\|d\|J\rangle}, \tag{5}$$

where $\langle J'\|d\|J\rangle$ is the reduced matrix element and $P_{F',F}$ are projectors on the excited and ground manifolds of magnetic sublevels. The second rank tensor, $\vec{\mathbf{T}} = \vec{\mathbf{f}} + i\vec{\mathbf{g}}$, describes the strength of the two-atom interaction as a function of atomic separation $r = |\mathbf{r}_1 - \mathbf{r}_2|$. The Hermitian part of the effective interaction Hamiltonian, $V_{dd}$, determines the dipole-dipole energy level shift, whereas the anti-Hermitian part, $\Gamma_{dd}$, gives rise to cooperative spontaneous emission, so that the total decay rate is given by the expectation value of $\Gamma_{tot} = \Gamma\left(\mathbf{D}_1^\dagger \cdot \mathbf{D}_1 + \mathbf{D}_2^\dagger \cdot \mathbf{D}_2\right) + \Gamma_{dd}$. In the near field, taking the limit $k_L r \to 0$, one finds $V_{dd} \Rightarrow \left(\mathbf{d}_1 \cdot \mathbf{d}_2 - 3(\hat{\mathbf{r}} \cdot \mathbf{d}_1)(\hat{\mathbf{r}} \cdot \mathbf{d}_2)\right)/r^3$, the quasi-static dipole-dipole interaction, and $\Gamma_{dd} \Rightarrow \Gamma\left(\mathbf{D}_1 \cdot \mathbf{D}_2^\dagger + \mathbf{D}_2 \cdot \mathbf{D}_1^\dagger\right)$, the Dicke super- (or sub) radiant interference term for in (or out of) phase dipoles. Because the level shift diverges for small $r$ whereas the cooperative emission remains finite, the time scales for coherent and incoherent interactions separate, providing a mechanism for controlled entanglement of the atoms, as discussed in Sec. I.

As a first estimate of the interaction matrix elements we will make a rather extreme simplifying approximation. We will assume that the strength of the excited dipole is *independent* of the internuclear coordinate separating the atoms. This will allow us to calculate the figure of merit analytically for some geometries. We will point out the range of validity of this approximation as we proceed and consider extensions in Sec. VI. Consider then a product state of two atoms, each described by a pure separable state of internal and external degrees of freedom $|\Psi\rangle = |\psi_{int}\rangle|\phi_{ext}\rangle$, where the internal state possesses a mean dipole moment vector oscillating along the spherical basis vector $\mathbf{e}_q$. The figure of merit for coherent dipole-dipole level shift follows from Eq. (4),

$$\kappa = \frac{\langle V_{dd}\rangle}{\langle \hbar\Gamma_{tot}\rangle} = \frac{-\hbar\Gamma\langle D_q^\dagger D_q\rangle_{int}\langle f_{qq}\rangle_{ext}}{2\hbar\Gamma\langle D_q^\dagger D_q\rangle_{int}\left(1+\langle g_{qq}\rangle_{ext}\right)} = \frac{-\langle f_{qq}\rangle_{ext}}{2\left(1+\langle g_{qq}\rangle_{ext}\right)}. \tag{6}$$

This factor depends only on *geometry*, the external states and the direction of polarization. It is independent of the strength of the dipole, since the same matrix element for the atoms' internal states appears both in the numerator and denominator. The average over the external state is carried out with respect to the relative coordinate probability density, having traced over the center of mass of the two-atom system.

It is necessary to devise interactions that minimize gate errors due to photon scattering and coherent coupling outside the logical basis. We will focus here on weak excitation of the dipoles. Adiabatic elimination of the excited states follows from second order perturbation theory in the limit of small saturation of the atomic transitions. When the detuning is large compared to the excited state level shifts, we can neglect the change of the level structure due to the dipole-dipole interaction, and consider saturation of the atomic levels, *independent of the external motional states*. For the case of alkali atoms the effective Hamiltonian on the ground state manifold is [30]

$$\begin{aligned}H_{dd} &= V_{dd} - i\frac{\hbar\Gamma_{dd}}{2} \\ &= -s\frac{\hbar\Gamma}{2}\sum_{q,q'=-1}^{1}\left(f_{qq'}+ig_{qq'}\right)\left(\left(\mathbf{D}_1\cdot\boldsymbol{\varepsilon}_L^*(\mathbf{r}_1)\right)\left(\mathbf{D}_1^\dagger\cdot\boldsymbol{\varepsilon}_q\right)\left(\mathbf{D}_2\cdot\boldsymbol{\varepsilon}_{q'}^*\right)\left(\mathbf{D}_2^\dagger\cdot\boldsymbol{\varepsilon}_L(\mathbf{r}_2)\right)+h.c.\right),\end{aligned} \tag{7}$$

with saturation $s \ll 1$. The coupling tensor is written here in the spherical basis. Physically, Eq. (7) represents a four photon process: absorption of a laser photon by one atom followed by coherent exchange of the excitation between the atoms via a virtual photon emission and absorption, and finally stimulated emission of a laser photon returning both atoms to the ground state. Because the virtual photon can be emitted in any direction, it is not an eigenstate of angular momentum with respect to the space-fixed quantization axis of the atoms. The quantum numbers $q$ and $q'$ represent two of the possible projections of its angular momentum on that axis.

We are left to consider the geometry of the trapping potentials and resulting external coordinate wave functions whose overlap with the dipole-dipole operator determines the level shifts. For deep traps we can approximate the motional states as harmonic oscillators. For the particular case of an isotropic trap, the spherical symmetry allows explicit evaluation of the interaction matrix elements. Consider two atoms in a common well, each described by a set of radial and angular momentum vibrational quantum numbers $|n,l,m\rangle$, with energy $E_{nl} = 2n + l + 3/2$, degeneracy $g_{nl} = (2n + l + 1)(2n + l + 2)/2$, and an internal state denoting one of the ground magnetic sublevels of a given hyperfine state $|F, M_F\rangle$. We have evaluated the matrix element,

$$\langle F, M_{F1}; n_1 l_1 m_1 | \otimes \langle F, M_{F2}; n_2 l_2 m_2 | V_{dd} | F, M'_{F1}; n'_1 l'_1 m'_1 \rangle \otimes | F, M'_{F2}; n'_2 l'_2 m'_2 \rangle, \tag{8}$$

in [13], from which we arrived at important selection rules. One finds that neither $M_{F1}, M_{F2}$, nor the total $M_{F1} + M_{F2}$ is a conserved quantity, but rather that $M'_{F1} + M'_{F2} = M_{F1} + M_{F2} + (q - q')$. The fact that these are not good quantum numbers can be seen immediately from the form of the interaction Hamiltonian, Eq. (4), which is neither a scalar with respect to rotations by hyperfine operators $\hat{\mathbf{F}}_1$, $\hat{\mathbf{F}}_2$, nor $\hat{\mathbf{F}}_1 + \hat{\mathbf{F}}_2$. Classically this is reflected in the fact that the dipole-dipole interaction is *not a central force*, and therefore the angular momentum of two classical dipoles about their center-of-mass is not a conserved quantity. Generally, internal angular momentum can be converted to rotational energy of the molecule if the atoms have multiply degenerate energy levels, obeying the selection rule, $m'_1 + m'_2 \pm (q - q') = m_1 + m_2$. In addition, the quantum numbers in Eq. (8) are restricted to conserve the total mechanical energy,

$$E_{tot} = E_1 + E_2 = E'_1 + E'_2 = E_{CM} + E_{rel} = E'_{CM} + E'_{rel} \tag{9a}$$

$$\Rightarrow n' = n + n'_1 + n'_2 - n_1 - n_2 + (l'_1 + l'_2 - l_1 - l_2 + l - l')/2. \tag{9b}$$

These selection rules impact the design of quantum logic gates. Besides decoherence, errors can occur due to coherent "leakage" to states outside the computational basis set. For example, without steps to suppress it, we can couple from two-qubit states within the logical basis to other degenerate, two-atom internal states through the exchange of virtual photons. Symmetry breaking must be employed through either magnetic fields and/or ac-stark shifts [13]. Leakage can also occur in the external quantum states for excited vibration modes of a common spherical well. For instance, the product state of two atoms, each with one quanta of vibration along *z*, can couple to the seven dimensional degenerate subspace of two quanta shared between the atoms,

$$|n_1 l_1 m_1\rangle \otimes |n_2 l_2 m_2\rangle = \{|010\rangle \otimes |010\rangle, |01-1\rangle \otimes |011\rangle, |011\rangle \otimes |01-1\rangle, \\ |020\rangle \otimes |000\rangle, |000\rangle \otimes |020\rangle, |100\rangle \otimes |000\rangle, |000\rangle \otimes |100\rangle\}. \tag{10}$$

We will consider an alternative protocol in Sec. IV employing excited vibrational states, but avoids this large degenerate subspace.

One final leakage channel we must address is coherent coupling into the excited state manifold. We must ensure that all population returns to the ground states after the logic gate is completed. One means to achieve this is to adiabatically connect the ground-manifold to the field-dressed levels. Adiabatic evolution is achieved when the level splitting between the two-atom ground and first excited eigenstates is sufficiently large compared to off-diagonal coupling caused by the changing catalysis excitation. An alternative approach is to work in the opposite limit, and apply sudden pulses, fast compared to $V_{dd}/\hbar$. A real excitation is then coherently exchanged between the atoms. A similar protocol in the form of a CNOT has been proposed by Lukin and Hemmer [31] for dipole-dipole interacting dopants in a solid state host.

## IV. Quantum Logic Gates

The quantum circuit model of quantum computing reduces the problem to precise operation of a universal set of quantum logic gates. Assuming reasonable fidelity in carrying out single qubit operations, the main challenge is to design and implement two-qubit gates. Though almost any entangling operation will give a universal set [32], we will consider two specific gates. A CPHASE gate can be implemented with the following protocol. If the catalysis field is tuned near the $|S_{1/2}, F_\uparrow\rangle \to |P_{3/2}\rangle$ resonance with detuning small compared to the ground-state hyperfine splitting, then nonnegligible dipoles are induced *only* for atoms in the logical-$|1\rangle$ states (Fig. 2a). If there are no off-diagonal matrix elements of the dipole-dipole interaction in the chosen logical

Page 12
*Special Issue of Forschritte der Physik*

basis, and assuming the gate is performed on a time scale much faster than the photon scattering rate, this causes only a non-zero *cooperative* level shift of the logical basis state $|1\rangle_+ \otimes |1\rangle_-$, and zero cooperative level shift of all other logical basis states (see Fig. 2b). Of course single atom light-shift interactions cause logical states to accumulate phase as well, but these can be removed through appropriate light-shift pulses acting on the separated individual atoms before and after they are made to interact. If the atomic pair is allowed to evolve in the presence of the catalysis field for a time $\tau = \hbar\pi / \langle V_{dd} \rangle$ we obtain $|1\rangle_+ \otimes |1\rangle_- \rightarrow -|1\rangle_+ \otimes |1\rangle_-$ with no change to the other logical basis states, as required for a CPHASE. The gate can be easily transformed into the familiar CNOT gate with single qubit Hadamard transformation before and after the cooperative phase [33]. A similar protocol can also be constructed to implement a $\sqrt{\text{SWAP}}$ gate [34]. Through an appropriate choice of logical basis and catalysis field, one can induce dipole-dipole couplings which are only off-diagonal in the logical basis, of the type $|1\rangle_+ \otimes |0\rangle_- \leftrightarrow |0\rangle_+ \otimes |1\rangle_-$. Applying the interaction for a time such that a $\pi/2$ rotation occurs in this subspace, we obtain the desired gate (Fig. 2c-d).

In the following we consider four examples which demonstrate the flexibility available for designing quantum logic gates. Except where otherwise noted, we will assign a logical basis set such as that given in Eq. (3). When the atoms are excited by $\pi$-polarized catalysis light, the figure of merit is given by Eq. (6), with $q = 0$, and the atoms are superradiant, $\Gamma_{tot} = 2\Gamma$. To complete our quantum logic protocol, we must choose the external coordinate wave function for our qubits to maximize the figure of merit.

### IV.A CPHASE Gate

Let us first consider the case of two atoms in the vibrational ground state sharing a common well. Though spherical wells maximize the radial overlap for atoms in their ground state, the dominant term in the interaction tensor is $f_{00} \sim Y_2^0(\theta)/(k_L r)^3$, which is orthogonal to the isotropic relative coordinate Gaussian wave function. This multipole component is nonzero, however, for nonspherical geometries and for higher motional states of the atoms in spherical wells.

One suitable design is to use ellipsoidal wells. Consider an axially symmetric harmonic potential with two atoms in the vibrational ground state, each described by a Gaussian wave packet with widths $\sigma_x = \sigma_y = x_0$ and $\sigma_z = z_0$. The figure of merit can be calculated numerically including radiation terms, as a function of $\eta_\perp = k_L x_0$ and $\eta_\parallel = k_L z_0$ as presented in [12]. In order to optimize the figure of merit, we consider an approximate analytic expression for $\kappa$ for tight localization. The external coordinate wave function separates into center of mass and relative coordinates, with RMS widths of the latter given by $\sigma_{x,rel} = \sqrt{2}x_0$ and $\sigma_{z,rel} = \sqrt{2}z_0$. Taking only the near field contribution to $H_{dd}$, where $\langle g_{00} \rangle_{ext} \approx 1$ and $\langle f_{00} \rangle_{ext} \approx -3 \langle P_2(\cos\theta)/(kr)^3 \rangle_{ext}$, we have from Eq. (6)

$$\kappa \approx -\frac{1}{4} \langle f_{00}(r,\theta) \rangle_{ext} = -\frac{3}{4} \int d^3 x |\psi_{rel}(r,\theta)|^2 \frac{P_2(\cos\theta)}{(k_L r)^3}$$
$$\frac{1}{16\sqrt{\pi}\eta_\perp^2 \eta_\parallel} \left[ -2 - 3\frac{\eta^2}{\eta_\perp^2} + 3\left(\frac{\eta^3}{\eta_\perp^3} + \frac{\eta}{\eta_\perp}\right) \tan^{-1}\left(\frac{\eta_\perp}{\eta}\right) \right], \quad (11)$$

where $\eta^{-2} = \eta_\parallel^{-2} - \eta_\perp^{-2}$. Keeping $\eta_\perp$ fixed while maximizing with respect to the ratio $\eta_\parallel/\eta_\perp$ gives $\kappa_{max} \approx -8.5 \times 10^{-3}/\eta_\perp^3$ for a ratio $(\eta_\parallel/\eta_\perp)_{max} \approx 2.18$. The relatively small prefactor can be attributed to two sources: the RMS width of the relative coordinate Gaussian wave function in three dimension is at least $\sqrt{6}$ times the RMS for a single particle in 1D, and the overlap of the angular distribution of the dipoles and $P_2(\cos\theta)$ is imperfect. As an example, given tight localizations $z_0 = \lambda/60$, $x_0 = \lambda/120$, corresponding to Lamb-Dicke parameters $\eta_\parallel = 0.1$, $\eta_\perp = 0.05$, we have $\kappa \approx -68$.

A disadvantage of using two atoms in a common ellipsoidal well is that the interaction potentials for different orientations of the relative coordinate destructively interfere with each other. For instance, for parallel dipoles aligned along $z$, $V_{dd} \sim -2d^2/r^3$ when the internuclear axis is along the polarization, and $V_{dd} \sim d^2/r^3$ for perpendicular separations. A possible solution is to use non-overlapping spherical wells, separated along $z$. We know that as this separation goes to zero, the dipole-dipole interaction goes to zero. We also know that $V_{dd} \sim 1/(kr)^3$ goes to zero as the separation goes to infinity. Thus at some intermediary value of atomic separation, the interaction must be maximum. For the case of two spherical wells separated by $\Delta z$, we can write the two particle external wave function as a product of isotropic ground state single particle Gaussians. The figure of merit follows as in Eq. (11),

$$\kappa = \left[ \frac{e^{-(\Delta \bar{z}/2)^2}}{\sqrt{\pi}} \left( \frac{1}{8} + \frac{3}{4\Delta\bar{z}^2} \right) - \frac{3 \operatorname{erf}(\Delta\bar{z}/2)}{4\Delta\bar{z}^3} \right] \frac{1}{\eta^3}, \quad (12)$$

where $\Delta\bar{z} = \Delta z/x_0$, and $\eta = k_L x_0$ with $x_0$ the single particle 1D localization. The form of Eq. (12) can be verified in two limits. For $\Delta z \gg x_0$, $\kappa \to -0.75(k_L \Delta z)^{-3}$, the expected figure of merit for two point dipoles separated by distance $\Delta z$, with dipole vectors aligned with the relative coordinate vector. For $\Delta z \ll x_0$, we find $\kappa \to -(\Delta \bar{z})^2/(80\sqrt{\pi}\eta^3)$, vanishing quadratically as the separation between wells goes to zero. A plot of $\kappa$ is shown in Fig. 3. The figure of merit is maximized at $\Delta z_{max}/x_0 \approx 2.5$ where $\kappa_{max} \approx -0.015/\eta^3$. For example, at $\eta = 0.05$, $\kappa_{max} \approx -123$. This is almost twice as good as the scheme using overlapping ellipsoidal wells with the same minimum localization.

Separated wells also have the advantage of reducing unwanted elastic and inelastic scattering processes which are significant for atomic separations on the order of a few Bohr radii. Given the relative coordinate wave function, the probability for two atoms to be separated by $r < a$ is

$$P(a) = \frac{1}{\Delta \bar{z} \sqrt{\pi}} \left( e^{-(\Delta \bar{z}+\bar{a})^2/4} - e^{-(\Delta \bar{z}-\bar{a})^2/4} \right) - \frac{1}{2} \left( \mathrm{erf}\left(\frac{\Delta \bar{z} - \bar{a}}{2}\right) - \mathrm{erf}\left(\frac{\Delta \bar{z} + \bar{a}}{2}\right) \right), \qquad (13)$$

where $\bar{a} \equiv a/x_0$. Proper design of the logic gate will require that within the radius of inelastic processes, this probability is sufficiently small. We will elaborate on this and other collisional issues in Sec. VI.

**IV.B $\sqrt{\mathrm{SWAP}}$ Gate**

Higher vibrational states of overlapping spherical wells can also be used to encode the qubit for controlled logic. For instance one quanta of vibration along $z$ in each atom could be considered to code for the logical-$|1\rangle$. This is ill suited as a logical basis, however, because of the problem of coherent leakage. The couplings given by the selection rules connect the logical basis to a seven dimensional degenerate subspace of two vibrational quanta shared between the atoms given in Eq. (10). Many of these couplings can be avoided, however, if instead we choose the so-called stretched states of vibration. Consider the logical basis

$$\begin{aligned} |1\rangle_\pm &= |F_\uparrow, M_F = \pm 1\rangle \otimes |n=0, l=1, m=1\rangle, \\ |0\rangle_\pm &= |F_\uparrow, M_F = \pm 1\rangle \otimes |n=0, l=0, m=0\rangle. \end{aligned} \qquad (14)$$

The logical-$|1\rangle$ states are circularly oscillating vibrational states. This induces nearly equal dipoles for atoms in logical-$|1\rangle$ and $|0\rangle$ states (see Fig. 2c).

Matrix elements of the dipole-dipole operator can then be calculated using Eq. (9). Unlike the previous cases discussed, the interaction operator is not diagonal in the computational basis set, $\{|0\rangle_+ \otimes |0\rangle_-, |0\rangle_+ \otimes |1\rangle_-, |1\rangle_+ \otimes |0\rangle_-, |1\rangle_+ \otimes |1\rangle_-\}$, but instead has the form [13]

$$V_{dd} = \frac{7}{4}\hbar\chi \begin{pmatrix} 0 & 0 & 0 & 0 \\ 0 & 1 & -1 & 0 \\ 0 & -1 & 1 & 0 \\ 0 & 0 & 0 & 4/7 \end{pmatrix}. \tag{15}$$

In addition, as dictated by the selection rule Eq. (9), the dipole-dipole interaction couples the logical basis states to a subspace of states with two shared quanta, $|n_1 m_1 l_1\rangle \otimes |n_2 m_2 l_2\rangle = \{|011\rangle \otimes |011\rangle, |022\rangle \otimes |000\rangle, |000\rangle \otimes |022\rangle\}$. The matrix elements are

$$\begin{aligned}
\langle 022| \otimes \langle 000|V_{dd}|011\rangle \otimes |011\rangle &= \langle 000| \otimes \langle 022|V_{dd}|011\rangle \otimes |011\rangle = -\hbar\chi/\sqrt{2} \\
\langle 022| \otimes \langle 000|V_{dd}|022\rangle \otimes |000\rangle &= \langle 000| \otimes \langle 022|V_{dd}|000\rangle \otimes |022\rangle = 9\hbar\chi/4 \\
\langle 022| \otimes \langle 000|V_{dd}|000\rangle \otimes |022\rangle &= -5\hbar\chi/4.
\end{aligned} \tag{16}$$

The couplings within the degenerate vibrational subspace of Eq. (16) describe an effective two-level system with coupling between the state $|011\rangle \otimes |011\rangle$ and the symmetric state $(|022\rangle \otimes |000\rangle + |000\rangle \otimes |022\rangle)/\sqrt{2}$. The antisymmetric state is uncoupled and "dark" to the interaction. The effective Rabi frequency within the coupled subspace is exactly $\chi$, thus there is a recurrence time $\tau = \pi/\chi$ for population in the vibrational state $|011\rangle \otimes |011\rangle$. For this interaction time the unitary operator in the logical basis is

$$U = \exp(-iV_{dd}\tau/\hbar) = \begin{pmatrix} 1 & 0 & 0 & 0 \\ 0 & \dfrac{e^{i\pi/4}}{\sqrt{2}}\begin{pmatrix} 1 & -i \\ -i & 1 \end{pmatrix} & 0 \\ 0 & & 0 \\ 0 & 0 & 0 & 1 \end{pmatrix} \tag{17}$$

This is the $\sqrt{\text{SWAP}}$ gate universal to quantum logic [?]. The figure of merit for this gate is

$$\kappa \approx -\frac{1}{4}\langle 1_+ 1_-|f_{00}|1_+ 1_-\rangle = -\frac{1}{140\sqrt{\pi}\,\eta^3} \approx -\frac{4.02 \times 10^{-3}}{\eta^3}, \tag{18}$$

which for $\eta = 0.05$ gives $\kappa \approx 32$. This figure could be improved by a factor of 3.5 if atoms oscillating along $z$ were used instead. However, an anisotropy would have to be introduced into the trapping potentials to suppress couplings to degenerate states outside the logical basis.

## V. Logic-Gate Error-Probability

### V.A. Optimizing the fidelity of a given protocol

For each of the quantum logic protocols examined in Sec. IV, the value of the figure of merit $\kappa$ implies an absolute lower limit on the logic-gate error-probability $\mathscr{L}$. This limit can be regarded as "fundamental", in the sense that it derives from a decoherence mechanism intrinsic to our scheme. In all cases we find a figure of merit $\kappa = c_\kappa \eta^{-3}$, as expected from the simple scaling law discussed in Sec. I, where the magnitude of the numerical constant $c_\kappa$ describes the "quality" of the protocol. With this knowledge it is possible to examine how the lower limit on $\mathscr{L}$ might in general be optimized through proper lattice design, and how it will scale as function of the most important lattice parameters. Even though it is unclear if we have at this time identified the most efficient protocols (see Sec. VI), it is instructive to estimate the $\mathscr{L}$ that might be achieved using the best $c_\kappa$ found in Sec. IV. Ignoring for the moment the possibility of inelastic atomic collisions, the gate fidelity is limited by spontaneous light scattering not just from the catalysis field, but also from the lattice field. The probability of scattering a catalysis photon during one operation of the gate can in principle be suppressed to an arbitrary degree (large $|\kappa|$) by tight localization of the wave packets. Tighter localization, however, requires a deeper lattice. Because there will always be a finite amount of laser power available to form the lattice this can be accomplished only at the cost of decreased lattice detuning, which in turn increases the probability of scattering a lattice photon during the gate operation. There exists then an optimum choice of lattice detuning, where the overall probability of error due to spontaneous light scattering is minimized.

We take the duration of a gate operation to be the time required to bring a pair of atoms together, perform the entangling operation, and return the atoms to their original positions. The catalysis field is present only during the entangling operation, of duration $\tau = \hbar\pi/|\langle V_{dd}\rangle|$. It then follows from the definition of the figure of merit, $\kappa = \langle V_{dd}\rangle/\langle \hbar\Gamma_{tot}\rangle$, that the probability of scattering a catalysis photon is $\mathscr{L}_C = 1 - \exp(-\Gamma_{tot}\tau) = 1 - \exp(-\pi/|\kappa|)$. The lattice can induce spontaneous photon scattering at any time throughout the duration of the gate, $T = t + \tau$, where $t$ is the time needed to transport the atoms together and apart again. The probability of scattering a lattice photon is then $\mathscr{L}_L = 1 - \exp(-\Gamma_L T)$, with $\Gamma_L$ the rate of scattering from the lattice. Because the lattice and catalysis detunings are very different we can treat scattering from the two fields as independent processes. The overall error-probability is then

$$\mathscr{L} = 1 - (1 - \mathscr{L}_C)(1 - \mathscr{L}_L) = 1 - \exp(-\pi/|\kappa|)\exp(-\Gamma_L T). \tag{19}$$

Whatever protocol is used, it seems likely that the time required to translate and superimpose atoms will be much longer than the duration of the cooperative interaction. We will express the total duration of the gate as $T \approx n 2\pi \omega_{osc}^{-1}$, where the vibrational oscillation frequency $\omega_{osc}$ sets the natural "clock speed" of the computer.

We can estimate the error probability for scattering a catalysis or lattice photon at a given intensity and detuning. Using the known relationships between intensity, well depth, wave packet extent and scattering rates [17], and assuming that the error rates are small, we find after some algebra that

$$\mathscr{L}_C \approx \frac{\pi}{|\kappa|} = \frac{\pi}{8 c_\kappa} \left( \frac{12 E_R}{\hbar \Gamma} \right)^{3/4} \left( \frac{I_1}{I_0} \right)^{-3/4} \left( \frac{\Delta_L}{\Gamma} \right)^{3/4}, \tag{20a}$$

$$\mathscr{L}_L \approx \Gamma_L T = \frac{\sqrt{3}\pi}{8} \left( \frac{2}{3} - \frac{1}{3 F_\uparrow} \right) \sqrt{\frac{\hbar \Gamma}{E_R}} n \left( \frac{I_1}{I_0} \right)^{1/2} \left( \frac{\Delta_L}{\Gamma} \right)^{-3/2}, \tag{20b}$$

where $I_1/I_0$ is the single-beam intensity in units of the saturation intensity, and $\Delta_L$ is the lattice detuning. In deriving Eq. (20b) we have assumed that an atom of e. g. the (+) species remains trapped at a node of the $\sigma_+$ lattice, and during the "transport phase" sees light from the $\sigma_-$ lattice at an average intensity roughly equal to half its value at the antinodes, making the $\sigma_-$ lattice by far the dominant source of photon scattering. Substituting Eqs. (20) into Eq. (19) and maximizing with respect to $\Delta_L$ yields a minimum error-probability of

$$\mathscr{P} \approx \pi \left[ \frac{n}{c_\kappa^2} \left( \frac{2}{3} - \frac{1}{3 F_\uparrow} \right) \frac{E_R}{\hbar \Gamma} \frac{I_0}{I_1} \right]^{1/3} \tag{21}$$

at a detuning of

$$\frac{\Delta_L}{\Gamma} = \frac{c_\kappa^{4/9}}{(12)^{1/9}} \left( \frac{2}{3} - \frac{1}{3 F_\uparrow} \right)^{4/9} \left( \frac{\hbar \Gamma}{E_R} \right)^{5/9} n^{4/9} \left( \frac{I_1}{I_0} \right)^{5/9}. \tag{22}$$

For example, for cesium with $F_\uparrow = 4$, $\hbar \Gamma / E_R = 2.5 \times 10^3$, excited with a large but not unrealistic intensity of $I_1 = 10^5 I_0$ (0.5 W of optical power in a beam with spotsize $w = 0.6$ mm), and for the protocol of optimally separated spherical wells, with $c_\kappa = 0.015$ and $n=2$ [15], we estimate a

Page 18
*Special Issue of Forschritte der Physik*

maximum error-probability of $\mathscr{P} \approx 0.08$ at an optimal detuning $\Delta_L \approx 6 \times 10^3 \Gamma$. This is sufficient to permit a non-trivial number of gate operations in initial experiments, and to produce entangled states with multiple atoms. Improvements in fidelity can be made with increased laser intensity, though from the scaling in Eq. (21) it is clear that very high-Q build-up cavities are required if substantial improvements are to be achieved. Ultimately, significant improvement of the fidelity will depend on improving both the scaling parameter $c_\kappa$ and the localization $\eta$, as we discuss in Sec. VI.

**V.B Measuring gate fidelity**

To develop better quantum logic protocols with higher fidelity it is crucial to obtain feedback from experiments. Fortunately this can be accomplished even with the sparse, randomly filled lattices routinely available in the laboratory today, without first solving the hard problems of lattice filling and single qubit addressing. Characterization of a quantum gate requires only *ensemble* measurements of the fidelity with which it achieves its truth table. In a sparse lattice this is complicated by the fact that only a small fraction of the qubits will be paired, i.e. have nearest neighbors on the lattice, compared to the much larger background of unpaired qubits. Nonetheless, it is possible to isolate the desired signal due to paired qubits from the background by first using the *gate itself* to identify the accidentally paired atoms, and then using radiation pressure to clear unpaired atoms from the lattice. Repeated application of the gate will then allow us to measure the error probability.

To emphasize this important point, we show here how to perform an ensemble measurement of the truth table of a CPHASE gate. In practice it is much easier to detect changes in atomic populations than phase shifts, so the CPHASE gate is converted to a CNOT in the standard fashion. For the purpose of illustration we will consider a specific input state $|1\rangle_+ \otimes |0\rangle_-$ for the CNOT gate, where the (+) and (−) species act as control and target bits, respectively. Only qubits with a nearest neighbor are brought together and acted upon by the CNOT operation; we will assume the gate leaves the majority of unpaired qubits behind in their original state. For the paired qubits, the gate succeeds with probability $1-\mathscr{P}$, flipping the target bit to the $|1\rangle_-$ state while leaving the control bit unchanged. There are three types of processes that can lead to gate errors: (i) the control or target qubit ends up in the correct state, but its partner is lost outside the computational basis; (ii) both qubits are lost; (iii) the control and/or target qubit ends up in the wrong logic state. For a $|1\rangle_+ \otimes |0\rangle_-$ input, correct operation of the gate leaves population in the $|1\rangle_+ \otimes |1\rangle_-$ state of the $F_\uparrow$ manifold, while errors populate the logical $|0\rangle_+$ and/or $|0\rangle_-$ states of the $F_\downarrow$ manifold, or states outside the computational basis. We can "flush" qubits in the $F_\downarrow$ manifold with near unit efficiency from the lattice with radiation pressure from a laser beam tuned to the $F_\downarrow \leftrightarrow F_\downarrow - 1$ cycling transition, transfer population of the $F_\uparrow$ manifold outside the logical basis to the $F_\downarrow$ manifold by Raman pulses, and flush once more. After we have flushed the failure modes of the gate in this fashion, all the failed gate operations have similar outcomes: one or both of the original, paired qubits are lost. The flushing also serves to remove all qubits of the target species which were originally unpaired.

The error probability can be determined through a repeated measurement. If we start out with $N$ pairs of qubits, the first CNOT/flushing cycle leaves $(1-\mathscr{P})N$ qubits of the target species in pairs. In addition $\alpha \mathscr{P} N$ new unpaired qubits of the target species have been created by gate errors that affect the control, but not the target, $\alpha$ being the fraction of such errors. If we rotate the target qubits back to the logical-zero state and carry out the CNOT/flushing cycle once more, there will be $(1-\mathscr{P})^2 N$ paired and $\alpha \mathscr{P}(1-\mathscr{P})N$ unpaired qubits of the target species. The success-probability $1-\mathscr{P}$ is then the fractional decrease in the number of target species qubits between the first and the second CNOT/flushing cycle. Similar sequences of CNOT/flushing cycles and single-qubit rotations allow to determine gate truth tables with other logical states as input. Even if a small fraction of the unpaired qubits are flipped by the CNOT gate, it is still possible to extract the error-probability, though it may require one or more additional gate operations to eliminate a sufficient number of the original unpaired atoms. It is also straightforward at any step to measure populations of hyperfine ground states outside the computational basis, in order to help distinguish between different failure mechanisms such as spontaneous emission and inelastic collision processes. Ultimately, we would need to measure the full process-fidelity of different quantum logic implementations [35], which is necessary to characterize fully how well each protocol preserves quantum coherence, by using control qubits in superpositions of the logical basis states. This requires a tomographic measurement of the entire density operator for the four-dimensional Hilbert space for each of 16 inputs, which can in principle be performed through single-qubit rotations and ensemble measurements of the populations in the two-qubit logical basis states, as demonstrated in NMR (see Cory in this volume).

## VI. Outlook

With recent advances in laser cooling and trapping technology, neutral atoms in optical lattices comprise an intriguing system for the exploration of quantum information science. In the short term development of quantum gate protocols can proceed using *ensemble measurements* on today's sparsely filled lattices, as discussed in Sec. V. In the longer term one must solve the dual challenges of filling the lattice in a controlled fashion, and addressing individual qubits during logic operations and readout. An interesting proposal by *Jaksch et al.* [36] have considered loading a Bose condensed atomic vapor into an optical lattice, showing that the system undergoes a Mott phase transition whereby the distribution of atoms might be frozen into uniform designer patterns. It is not yet known if this or any other preparation method will prove reliable enough to permit large scale "computations". It also remains to develop methods whereby individual qubits can be addressed for manipulation and readout. Direct spatially resolved imaging will likely prove impossible in standard lattices, where individual atoms are spaced by half an optical wavelength. It may be possible to separate qubits further, e. g. by filling the lattice with a sparse but regular array of atoms or by using $CO_2$ laser light for trapping [11], and to increase the spatial resolution by using a near-UV transition for the Raman and catalysis lasers. One can also envision tomographic methods akin to Magnetic Resonance Imaging wherein the Zeeman shift from a non-uniform magnetic field is used to correlate qubit position and frequency, thus mapping coordinate space into the spectral domain. Considering for example atomic Cs, a magnetic field gradient of order 100 Gauss/cm would lead to a barely resolvable differential Zeeman shift between magnetic sublevels of order 10 KHz per wavelength. These and other solutions to the addressing problem must all be explored in more detail. Also, one should carefully consider what kinds of information processing tasks, might be performed with quantum cellular automata [37] or special codings where one has only limited local control over the qubits [38].

Even if these technical challenges can be met, it remains to be seen whether this system can meet the stringent requirements for universal quantum computing. The error-probability of the two-qubit logic gate we have considered, finite due to spontaneous emission, is sufficient to allow us to create entangled-states of multiple atoms. This holds promise for a variety of applications in precision measurement [7] and quantum simulations [39]. Such applications are possible in the short term, based on ensemble preparation and measurement on the sparse, randomly filled lattices that are available in today's laboratory experiments. In the much longer term, the promise of universal fault-tolerant quantum computing places very strong constraints on the physical system. Specifically, fault-tolerant quantum computation demands an extremely low error-probability, e.g. $\mathscr{P} < 10^{-4}$ [40]. Particularly intriguing is the prospect for exploiting the intrinsic geometry for error correction. Assuming control to move atoms along any of the three orthogonal directions, blocks of qubits may be coded and then made to interact *in parallel* as required for fault-tolerant processing [40]. An example of such coding has been considered by Briegel *et al.* [15]. Fortunately we see possibilities for substantial improvement in gate-fidelity by extending the theoretical analysis beyond the simplifying assumptions considered up to this point.

Given the scaling arguments of Sec. I, the dipole-dipole figure of merit has the form $\kappa \sim c_\kappa \eta^{-3}$. One may consider the ultimate limits on these parameters. The localization $\eta$ depends on the quality of the trap and possible implementations of wave packet control, such as squeezing [27]. The parameter $c_\kappa$ is specific to the details of the protocol and the approximations of our analysis. Under the assumptions of our model, $c_\kappa$ is determined solely by geometry and might also benefit from wave packet engineering to maximize the relevant dipole-dipole multipole component of the relative coordinate probability distribution. More importantly, we must address the limitations of our model. Implicitly we have assumed that the strength of the excited dipoles is *independent* of the relative coordinate between the dipoles. This approximation is valid when the detuning of the catalysis laser is very large compared to the dipole-dipole energy level shifts of the excited state over the range of internuclear distances supported by the relative coordinate probability density. However, this approximation will break down for smaller detunings, especially for closely overlapping wave packets where the catalysis and/or trapping fields will be resonant with the excited manifold for some internuclear separations (the "Condon points"). Stated in other terms, when the internuclear separation becomes smaller that the optical wavelength one must consider the *molecular* rather than *atomic* energy level structure of the excited states.

Extending our model to include molecular resonance may have substantial impact on the dipole-dipole figure of merit. For example, we have argued that for atoms in the vibrational ground state of a common isotropic spherical well, the dipole-dipole coherent interaction is zero due to destructive interference when integrating over all angles of the relative coordinate vector. However, at finite detuning, e.g. red of atomic resonance, the catalysis field will preferentially excite the attractive potential, leading to finite interaction, and an increase in the parameter $c_\kappa$. Another example is the use of subradiant states. We have implicitly assumed that our dipoles are excited in phase, leading to Dicke superradiance. However, molecular resonances exist for dipoles oscillating out of phase, which might be excited with a sufficiently intense catalysis field. This too would impact the maximum possible value of $c_\kappa$ in the figure of merit.

Besides providing a more accurate model for calculating the interaction strength, inclusion of the molecular potentials is crucial to account for possible inelastic collisional processes [41] which have been neglected in the basic treatment. If Condon points exist, atoms will likely photo-associate, accelerate rapidly on the excited potential curves and be ejected from the trap. Less catastrophic, but detrimental to the operating fidelity of the quantum logic gate, is "leakage" of population outside the computational basis set through spin couplings or a strong perturbation of the motional states. These phenomena must be modeled if we are to assess the capability of the logic gate. Careful choice of the parameters may avoid these resonances over the extent of the relative coordinate probability density.

Viewing this issue from the reverse perspective, molecular resonance can act as a *resource* for designing logic gates. A possible route to coherent two-qubit interactions is to make use of *bound* molecular states. In the mechanisms discussed above, atoms in a particular logical basis state are connected adiabatically to a dressed-state, and thereby accumulate a cooperative phase shift. Alternatively, one can work in the opposite limit and rapidly induce a coherent Rabi $2\pi$-rotation between the target two-bit and an auxiliary molecular resonance and thereby accumulate a conditional $\pi$–phase shift. For example, through application of a magnetic field, a bound-state of two atoms associated with one hyperfine state can be made resonant with a scattering state associated with the other. Coherent control of this so called Feshbach free-bound transitions has recently been studied by Mies *et al.* [42]. Another option is to use an excited bound molecular-state as the auxiliary two-atom state. A particularly intriguing choice is a bound-state of the so-called "purely long-range" potentials observed in photo-association spectroscopy in ultra-cold collisions [43]. These states have the unique property that their inner turning point is well beyond the range of most chemical binding processes where inelastic collisions (such as fine-structure changing collisions) dominate

Whichever protocol ultimately holds the greatest promise, the rich structure of the neutral-atom/optical-lattice system provides new avenues for explorations of quantum control and information processing. The potential payoff in new understandings of the foundations of quantum mechanics, improvements in precision measurement, novel collision physics, and the promise of quantum computing makes this an exciting arena for new interdisciplinary research.

We gratefully acknowledge helpful discussions with Carlton Caves, Paul Alsing, John Grondalski and Shohini Ghose, and Gary Herling. IHD and GKB acknowledge support from the National Science Foundation (Grant No. PHY-9732456). PSJ was supported by the NSF (Grant No. PHY-9503259), by the Army Research Office (Grant No. DAAG559710165), and by the Joint Services Optics Program (Grant No. DAAG559710116).

**FIGURE CAPTIONS**

**FIG. 1.** Schematic of a 3D blue-detuned optical lattice. (a) Two pairs of $\pi$-polarized beams provide transverse confinement, and the beams along $z$ in the lin-θ-lin configuration provide longitudinal confinement in $\sigma_+$ and $\sigma_-$ standing waves. (b) Potential surfaces for the $(\pm)$-atomic species, described in the text, shown here as in gray and white, are moved along the $z$-axis through a rotation of the angle $\theta$ between polarization vectors.

**FIG. 2.** Energy level structure of the logical basis associated with two-qubit logic gates. Basis states are denoted for $(\pm)$-species as described in the text. (a,b) CPHASE configuration: The "catalysis field" excites dipoles only in the logical-$|1\rangle$ states, chosen for both species in the upper ground hyperfine manifold $F_\uparrow$. The dipole-dipole interaction is diagonal in this basis and results solely in a level shift of the $|1\rangle_+ \otimes |1\rangle_-$ state. Operation of this gate with high fidelity requires this shift to be large compared to the cooperative linewidth. (c,d) $\sqrt{\text{SWAP}}$ configuration. The logical basis is encoded in the vibrational degree of freedom as described in Sec. V.B. For an appropriate choice of geometry and pulse timing, there is an off-diagonal coupling between the logical states $|1\rangle_+ \otimes |0\rangle_-$ and $|0\rangle_+ \otimes |1\rangle_-$, yielding a $\sqrt{\text{SWAP}}$.

**FIG. 3.** Dipole-dipole figure of merit $\kappa$ for spherically symmetric Gaussian wave packets with width $x_0$, normalized to the Lamb-Dicke parameter $\eta = kx_0$, as a function of the normalized separation $\Delta \bar{z} = \Delta z / x_0$. Maximum $|\kappa| \approx 0.015 / \eta^3$ is achieved at $\Delta \bar{z} \approx 2.5$.

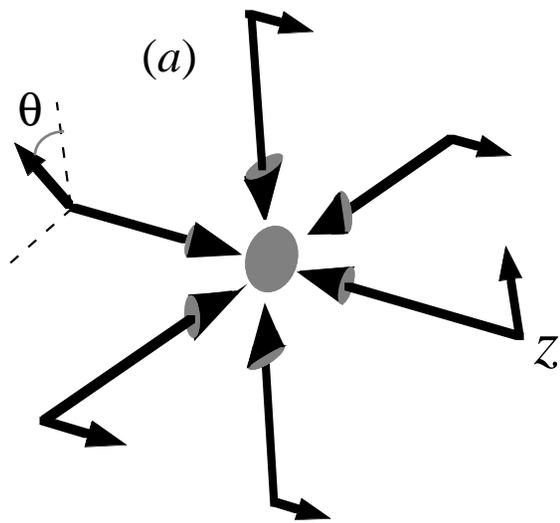 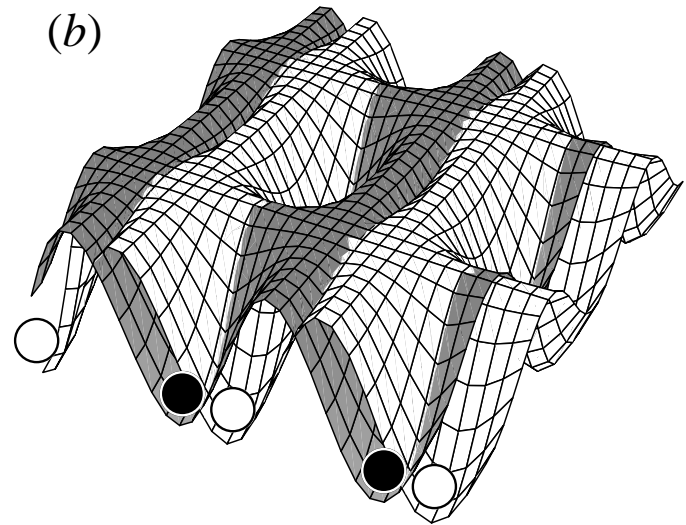

*Deutsch et al.* Fig. 1

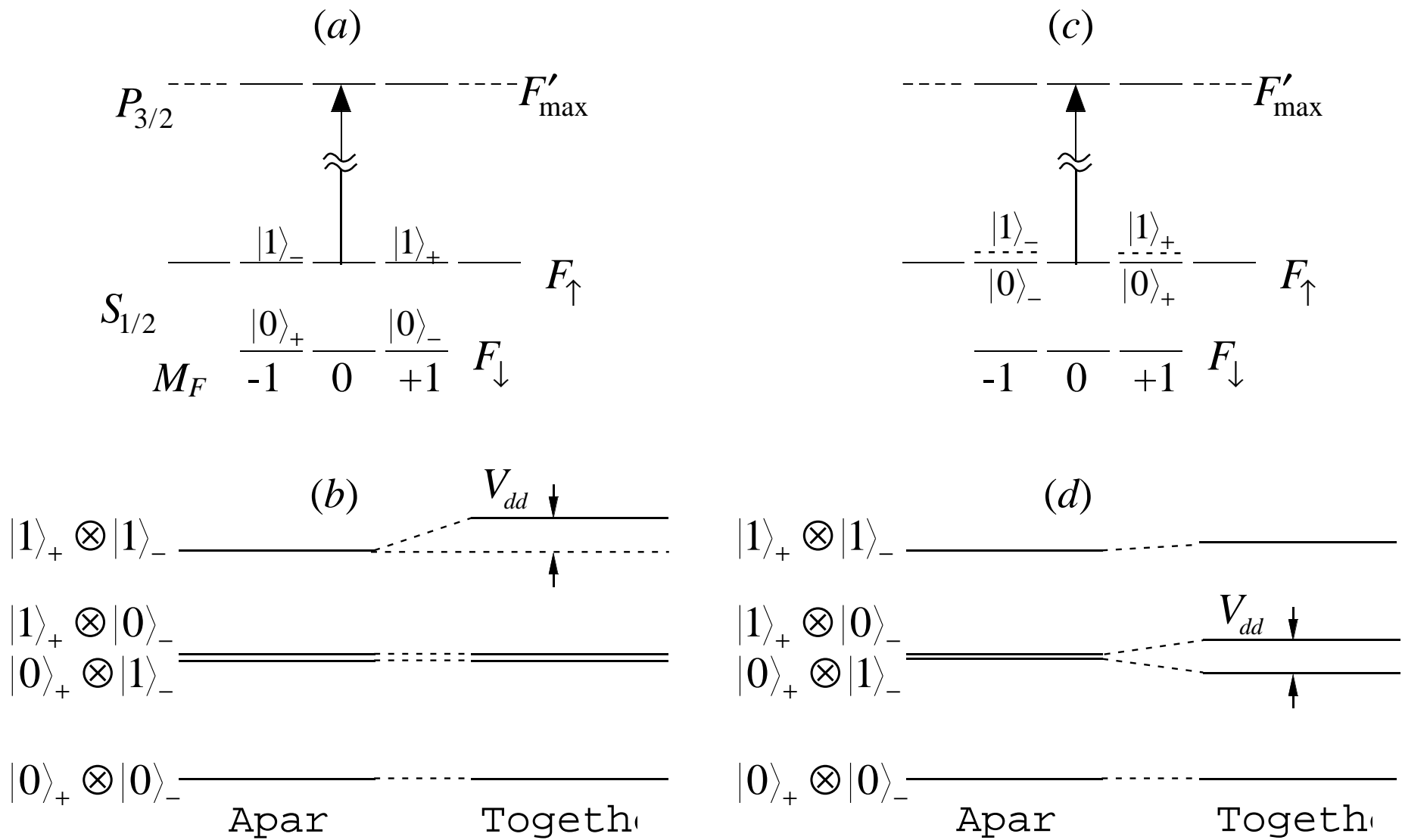

Deutsch et al. Fig. 2

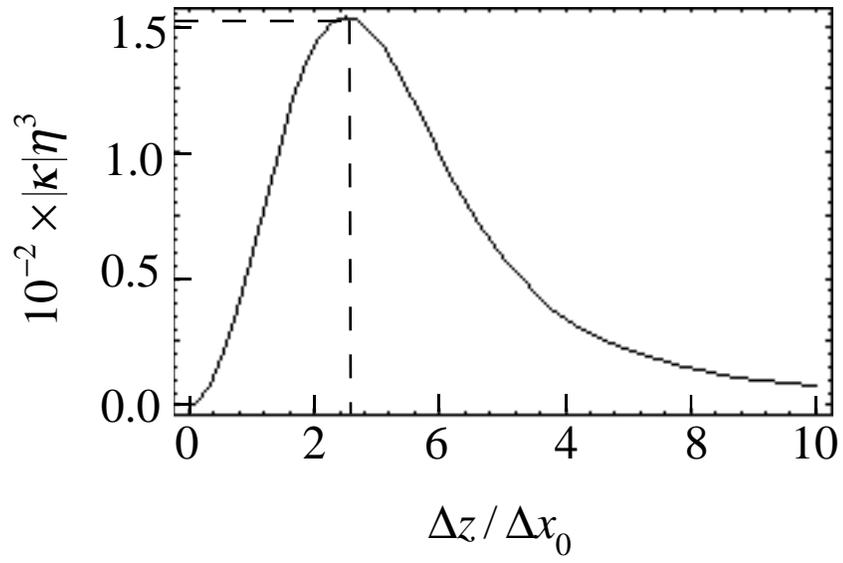

*Deutsch et al.* Fig. 3